\documentclass[aps,prd,twocolumn,groupedaddress,showpacs,preprintnumbers,draft]
{revtex4}
\usepackage{epsf} 
\begin{document}
\preprint{BROWN-HET-1391}
\title{Note on the Robustness of the Neutrino Mass Bounds from Cosmology}
\author{Robert H. Brandenberger}%
	\email[Email address :]{rhb@het.brown.edu}
\affiliation{Department of Physics, Brown University, Providence, Rhode
Island 02912, USA and\\
Perimeter Institute for Theoretical Physics, Waterloo, Canada N2J 2W9} 
\author{Anupam Mazumdar}%
     \email[Email address :]{anupamm@hep.physics.mcgill.ca}
\affiliation{Department of Physics, McGill University, Montr\'eal, Qu\'ebec, 
Canada H3A 2T8}
\author{Masahide Yamaguchi}%
	\email[Email address :]{gucci@het.brown.edu}
\affiliation{Department of Physics, Brown University, Providence, Rhode
Island 02912, USA}
\begin{abstract}
The recent high precision maps of cosmic microwave anisotropies combined
with measurements of the galaxy power spectrum from new large-scale
redshift surveys have allowed stringent bounds on the sum of the 
neutrino masses to be placed. The past analyses, however, have
implicitly assumed that the spectrum of primordial density fluctuations
is adiabatic and coherent, as predicted in the simplest models of
inflation. In this paper, we show that the limits hold even if the
assumption on the primordial power spectrum is relaxed to allow
for a contribution of nonadiabatic, incoherent fluctuations such
as would be predicted by topological defects. 
\end{abstract}
\maketitle
\def\Box{\nabla^2}  
\def\ie{{\em i.e.\/}}  
\def\eg{{\em e.g.\/}}  
\def\etc{{\em etc.\/}}  
\def\etal{{\em et al.\/}}  
\def\S{{\mathcal S}}  
\def\I{{\mathcal I}}  
\def\mL{{\mathcal L}}  
\def\H{{\mathcal H}}  
\def\M{{\mathcal M}}  
\def\N{{\mathcal N}} 
\def\O{{\mathcal O}} 
\def\cP{{\mathcal P}} 
\def\R{{\mathcal R}}  
\def\K{{\mathcal K}}  
\def\W{{\mathcal W}} 
\def\mM{{\mathcal M}} 
\def\mJ{{\mathcal J}} 
\def\mP{{\mathbf P}} 
\def\mT{{\mathbf T}} 
\def\mR{{\mathbf R}}
\def\mS{{\mathbf S}}
\def\mX{{\mathbf X}}
\def\mZ{{\mathbf Z}}
\def\eff{{\mathrm{eff}}}  
\def\Newton{{\mathrm{Newton}}}  
\def\bulk{{\mathrm{bulk}}}  
\def\brane{{\mathrm{brane}}}  
\def\matter{{\mathrm{matter}}}  
\def\tr{{\mathrm{tr}}}  
\def\normal{{\mathrm{normal}}}  
\def\implies{\Rightarrow}  
\def\half{{1\over2}}  
\newcommand{\da}{\dot{a}}
\newcommand{\db}{\dot{b}}
\newcommand{\dn}{\dot{n}}
\newcommand{\dda}{\ddot{a}}
\newcommand{\ddb}{\ddot{b}}
\newcommand{\ddn}{\ddot{n}}
\newcommand{\ba}{\begin{array}}
\newcommand{\ea}{\end{array}}
\def\be{\begin{equation}}
\def\ee{\end{equation}}
\def\bea{\begin{eqnarray}}
\def\eea{\end{eqnarray}}
\def\bs{\begin{subequations}}
\def\es{\end{subequations}}
\def\g{\gamma}
\def\G{\Gamma}
\def\vp{\varphi}
\def\mpl{M_{\rm P}}
\def\ms{M_{\rm s}}
\def\ls{\ell_{\rm s}}
\def\lp{\ell_{\rm pl}}
\def\l{\lambda}
\def\gs{g_{\rm s}}
\def\d{\partial}
\def\co{{\cal O}}
\def\sp{\;\;\;,\;\;\;}
\def\spa{\;\;\;}
\def\r{\rho}
\def\dr{\dot r}
\def\dt{\dot\varphi}
\def\e{\epsilon}
\def\k{\kappa}
\def\m{\mu}
\def\n{\nu}
\def\om{\omega}
\def\tn{\tilde \nu}
\def\p{\phi}
\def\vp{\varphi}
\def\P{\Phi}
\def\r{\rho}
\def\s{\sigma}
\def\t{\tau}
\def\x{\chi}
\def\z{\zeta}
\def\a{\alpha}
\def\b{\beta}
\def\de{\delta}
\def\bra#1{\left\langle #1\right|}
\def\ket#1{\left| #1\right\rangle}
\newcommand{\stt}{\small\tt}
\renewcommand{\theequation}{\arabic{section}.\arabic{equation}}
\newcommand{\eq}[1]{equation~(\ref{#1})}
\newcommand{\eqs}[2]{equations~(\ref{#1}) and~(\ref{#2})}
\newcommand{\eqto}[2]{equations~(\ref{#1}) to~(\ref{#2})}
\newcommand{\fig}[1]{Fig.~(\ref{#1})}
\newcommand{\figs}[2]{Figs.~(\ref{#1}) and~(\ref{#2})}
\newcommand{\GeV}{\mbox{GeV}}
\newcommand{\Mpc}{\mbox{Mpc}}
\def\ricci{R_{\m\n} R^{\m\n}}
\def\riemann{R_{\m\n\l\s} R^{\m\n\l\s}}
\def\triemann{\tilde R_{\m\n\l\s} \tilde R^{\m\n\l\s}}
\def\tricci{\tilde R_{\m\n} \tilde R^{\m\n}}
\section{Introduction}

Recently, there has been a lot of progress in neutrino physics. Measurements
of solar and atmospheric neutrino fluxes have yielded tight bounds on
the mass differences between the different neutrino mass eigenstates
(see e.g. \cite{rev} for recent reviews). The bounds on the absolute
masses from direct measurements, however, are much weaker. Currently, the
bounds on the {\it absolute} neutrino masses coming from indirect
cosmological considerations are much stronger. In particular, given
the recent high precision data on cosmic microwave background (CMB) 
anisotropies and the increased information about the galaxy
power spectrum based on the recently completed large-scale
galaxy redshift surveys,  it is possible to deduce tight bounds 
on the sum of the neutrino masses \cite{list}. These analyses, however,
make use of certain assumptions about the evolution of the early Universe.
Specifically, they make use of theoretical predictions for the power
spectra of the galaxy distribution, derived under the assumption that
the primordial spectrum of cosmological perturbations was purely
adiabatic and coherent, as predicted in simple single field models
of inflation. In this paper we show that the bounds derived in \cite{list}
are {\it robust} against the addition of a contribution to the primordial
power spectrum which is nonadiabatic and incoherent, namely a
contribution coming from a distribution of topological defects.

The cosmological limits on the sum of the neutrino masses are derived by
combining the high precision data of CMB anisotropies with the
measurements of the galaxy power spectrum which have recently improved
in accuracy upon completion of the 2dF (two degree field) large scale
structure survey. Given the assumption that the primordial power
spectrum is purely adiabatic and coherent (as predicted in simple single
scalar field models of inflation), the CMB angular power spectrum
determines to good accuracy the slope of the primordial power spectrum
and the cosmological parameters (see e.g. \cite{parsrev} for recent
reviews). The most sensitive dependence on the neutrino masses comes
from the smallest scales, those which are not (yet) probed by CMB
anisotropies but rather by the galaxy power spectrum. The reason for
this dependence lies in the phenomenon of {\it neutrino free streaming}
\cite{nfs}. Because of their small mass, the neutrinos have large
velocities at $t_{eq}$, the time of equal matter and radiation.  Hence,
neutrino density perturbations on comoving scales smaller than the
neutrino free streaming length (the distance the neutrinos travel in one
Hubble expansion time) at $t_{eq}$ are suppressed (the suppression
factor increases exponentially as a function of the wavenumber). Hence,
if the contribution of the neutrinos to $\Omega$ (the energy density in
units of the critical density) is too large, there will be insufficient
power to explain the observed magnitude of the galaxy power spectrum. A
lower bound on the galaxy power spectrum thus translates into an upper
bound on the contribution of neutrinos to $\Omega$, and thus to an upper
bound on the sum of the neutrino masses. Note that the most stringent
constraints come from the smallest length scales for which the galaxy
power spectrum can be reliably determined.

However, the assumption that the primordial spectrum of fluctuations
is purely adiabatic and coherent is a very restrictive one. Even in
the context of scalar field driven inflationary cosmology, as soon as
one considers models with more than one scalar field, it is possible
to obtain a contribution of entropy fluctuations \cite{LindeMukh}.
Many inflationary models based on grand unified theories 
(see e.g. \cite{Rachel}) or on the brane inflation scenario (see 
e.g. \cite{Tye}) predict a contribution to the power spectrum from 
cosmic strings, yielding a contribution of isocurvature and
incoherent fluctuations. It has been known for a long time \cite{BKST,BKT}
that the transfer function which relates the primordial power spectrum
to the present power spectrum changes dramatically on small distance
scales if a contribution of seed perturbations such as those
created by cosmic strings is added. The basic point is that the
cosmic strings constitute density fluctuations seeds which are not
erased by neutrino free streaming. Thus, in a seed model, the accretion
of neutrinos on small scales is delayed (and thus reduced in amplitude)
but not prevented. Hence, it appears at first sight that even a small
addition of cosmic string seed fluctuations to the primordial power
spectrum might dramatically loosen the cosmological bounds on the
sum of the neutrino masses. Here, we shall demonstrate that this is
in fact {\it not} the case.

The outline of our analysis is as follows. We will first consider the
largest contribution of cosmic strings to the angular power spectrum on
large angular scales which is consistent with the current data on CMB
anisotropies, making use of the recent Wilkinson Microwave Anisotropy
Probe (WMAP) results \cite{WMAP}.  This result will then determine the
contribution of the cosmic strings to the mass power spectrum. In this
context, it is important to take into account the fact that the
perturbations produced by the strings are isocurvature in nature (as
emphasized e.g. in \cite{TTB}).  This will effect the ratio of their
contributions to the angular CMB power spectrum and the matter power
spectrum. We then use the (approximately) known transfer function for
strings and cold dark matter (which is larger on small scales than the
standard transfer function for adiabatic fluctuations with cold dark
matter) to estimate the total matter power spectrum on small scales, and
analyze whether the loss of power on small scales when increasing the
neutrino contribution to $\Omega$ can be compensated by the increase in
power coming from the presence of cosmic string seed perturbations.

\vskip.5cm
\section{Analysis}

The recent WMAP data on the angular power spectrum of CMB anisotropies
has mapped out with high precision the region of the power spectrum
corresponding to the first acoustic peak. This peak is narrowly
centered at a value $l = 220 \pm 1$ \cite{WMAP}, in good agreement with a 
cosmology in which the Universe is spatially flat and the primordial
spectrum of fluctuations is coherent and adiabatic. Cosmic strings,
on the other hand, give rise to isocurvature fluctuations which
are incoherent. As a consequence, there are no marked acoustic
oscillations in the angular power spectrum \cite{PST,Albrecht,Battye,Allen},
but only a fairly broad Doppler peak \cite{Periv}. Thus, the present
data tightly constrain the maximal contribution of strings to the
angular power spectrum of CMB anisotropies on large angular scales.
 
Since the theoretical predictions for the spectrum of CMB
anisotropies resulting from cosmic strings in the acoustic peak region 
are quite uncertain, the bounds on the contribution of cosmic strings
to the CMB angular power spectrum on large scales are also not
certain. Different investigations have yielded bounds between
$1 \%$ \cite{Avelino} and $10 \%$ \cite{Pogosian} (see also
\cite{Endo}). We will denote the upper bound as $f$.  

We will assume that the fraction $f_{CDM}$ of the dark matter is cold,
and the rest hot. There are two sources of fluctuations, the
conventional adiabatic perturbations (e.g. produced by quantum
fluctuations during inflation) and the cosmic strings. We will assume
that the cosmic strings are not correlated with the inflationary
perturbations. In this case, the cross terms (between the
string fluctuations and the adiabatic perturbations) to the angular CMB
power spectrum and to the mass power spectrum vanish, and the mass power
spectrum $P(k)$ calculated in linear cosmological perturbation theory
can be written as
\begin{equation} \label{basic}
P(k) \, = \, (1 - f) P_a(k) + f P_{CS}(k) \, ,
\end{equation}
where $P_a$ and $P_{CS}$ are the power spectra for a pure adiabatic
model and a pure cosmic string model, respectively.

Since the tightest constraints on the fraction of hot dark matter come
from the observed power spectrum of matter on the smallest scales
for which linear perturbation theory is adequate and for which 
observational results are robust, we
will in the following focus on these scales. For these scales, the
contribution of hot dark matter to the matter power spectrum for
adiabatic fluctuations is negligible, and thus
\begin{equation} \label{adspec}
P_a(k) \, \simeq \, f_{CDM}^2 P_0(k) \, ,
\end{equation}
where $P_0$ is the power spectrum of the current concordance model
(scale invariant adiabatic fluctuations in the $\Lambda$CDM model
\footnote{We will consider the same background cosmological parameters
$\Lambda$, $h$, etc., as in this concordance model. As emphasized in
\cite{Endo}, the addition of cosmic strings to the model may yield a
slightly different best fit value for the parameters. We will return to
this issue in future work.}). The result (\ref{adspec}) is not exact for
two reasons: first, since cold dark matter does cluster about the
primordial perturbations, neutrino free streaming delays but does not
totally prevent the clustering of neutrinos. This effect would lead to a
spectrum larger than given in Eq. (\ref{adspec}). On the other hand, the
local clustering of the cold dark matter occurs as if the Universe were
slightly open. This effect tends to reduce the spectrum from the result
(\ref{adspec}). However, both of these effects are small if the fraction
of hot dark matter is not too large. Even if we use the exact transfer
function of adiabatic fluctuations \cite{WDM}, the final result does not
change significantly.

On small scales, the transfer function for a model with cosmic strings
and cold dark matter is different from the transfer function in the
concordance model \cite{AS1}. The transfer function is, in fact, much
larger. Thus, as noted in \cite{AS1}, a pure cosmic string model
which is $\sigma_8$-normalized (i.e. normalized such that the power
spectrum on a length scale of $8 h^{-1}$Mpc agrees with observations)
produces too much small scale structure. Hot dark matter clustering
onto cosmic string loops, on the other hand, is delayed but not
prevented. Thus, the hot dark matter power spectrum in a cosmic string
model on small scales is not exponentially suppressed. However, the
hot dark matter cosmic string power spectrum on small scales is much
smaller than the cold dark matter power spectrum, so that to a good
approximation we can use
\begin{equation} \label{csspec}
P_{CS}(k) \, \simeq \, f_{CDM}^2 P_{(0, CS)} \, ,
\end{equation}
where $P_{(0, CS)}$ denotes the power spectrum in a pure cosmic string
model with cold dark matter and the current concordance parameters.

The question we would like to ask is whether the contribution of cosmic
strings to the power spectrum, given by Eqs. (\ref{basic}) and
(\ref{csspec}) is large enough to compensate for the loss in small scale
power from the adiabatic fluctuations [given by Eq. (\ref{adspec})] if
$f_{CDM} < 1$ and to render the theory compatible with
observations. More specifically, we are interested in finding the
smallest value of the cold dark matter fraction $f_{CDM}$ for which the
theoretical power spectrum $P(k)$ exceeds the observational result
$P_{obs}(k)$, i.e.
\begin{equation} \label{ineq}
P(k) \, > \, P_{obs}(k) \,
\end{equation}
on the smallest scales for which high quality observational results
exist and for which we can trust the results of linearized perturbation
analysis, namely $k = 0.2 h \Mpc^{-1}$ \cite{Hannestad}.. 

The results of \cite{AS2} indicate that on scales of $k = 0.2 h
\Mpc^{-1}$ the cosmic string cold dark matter power spectrum $P_{(0,
CS)}$ does not exceed the adiabatic cold dark matter spectrum $P_0(k)$
by more than a factor of 2. In fact, even on scales as low as $1
h^{-1}$Mpc, the difference is not more than a factor of 10. The results
of \cite{AS2} are for models without a cosmological constant. However,
as demonstrated in \cite{Wu}, on the smaller scales of relevance here,
the matter power spectrum does not depend on the value of the
cosmological constant in a significant way, and thus we can make use of
the results of \cite{AS2}. On the other hand, it is well known that the
perturbations seeded by strings are isocurvature in nature (see
e.g. \cite{TTB}). For isocurvature fluctuations, the CMB temperature
anisotropies on large scales are given by
\begin{equation}
{{\delta T} \over T} \, = \, 2 \Phi \, ,
\end{equation}
where $\Phi$ is the relativistic gravitational potential (see e.g.
\cite{MFB}) which determines the magnitude of the mass perturbations.
This is to be compared with
\begin{equation}
{{\delta T} \over T} \, = \, {1 \over 3} \Phi \, ,
\end{equation}
in the case of adiabatic fluctuations. Thus, if we normalize the
power spectrum by the CMB anisotropies on large angular scales,
the cosmic string power spectrum on large scales is a factor
of 36 smaller than the power spectrum for adiabatic fluctuations.
Hence, we see that on scales relevant to the large-scale structure,
the decrease in the matter power spectrum which results from the
addition of strings as a secondary source of cosmological
fluctuations far exceeds the increase which is obtained by the
change in the transfer function. Using the optimistic value 2
for the increase in the power due to the change in the transfer
function on the relevant scale of $k = 0.2 h \Mpc^{-1}$, we
conclude that in a model with both cosmic string and adiabatic
scale-invariant fluctuations
\begin{eqnarray} \label{ineq2}
P(k) \, &=& \, (1 - f) P_a(k) + f P_{CS}(k) \nonumber \\
        &\leq& (1 - f) f_{CDM}^2 P_0(k) + f f_{CDM}^2 {2 \over {36}} P_0(k) 
                       \nonumber \\
        &<& f_{CDM}^2 P_0(k) \, .
\end{eqnarray}

In a model with only adiabatic perturbations, the 
upper bound on the sum of the neutrino masses obtained
by combining the large-scale structure and CMB data comes from
demanding that
\begin{equation}
P(k) \, \geq \, P_l(k) \, ,
\end{equation}
where $P_l(k)$ is the observational lower bound on the
mass spectrum. Since in this case 
\begin{equation}
P(k) \, \simeq \, f_{CDM}^2 P_0(k) \, ,
\end{equation}
and knowing that the $P_0(k)$ fits the date well,
this gives a lower bound on $f_{CDM}$ which is equivalent
to an upper limit on the sum of the neutrino masses. 

Combining these last three equations, we conclude that the inclusion
of a contribution of cosmic strings to the primordial power
spectrum does not lead to a relaxation of the cosmological bounds
on the sum of the neutrino masses.

\vskip.5cm
\section{Discussion}

In this paper we have shown that the current cosmological neutrino mass
limits, which were derived under the assumption that the spectrum of
cosmological perturbations is purely adiabatic, are robust against the
addition of the maximal contribution which cosmic strings could make to
the primordial power spectrum. {\it A priori}, this is not obvious,
since cosmic string seeds survive neutrino free streaming and,
therefore, clustering on small scales (scales which yield the tightest
constraints on the neutrino masses) is not prevented but only
delayed. However, if the power spectrum is normalized by the large-scale
CMB anisotropies, then the contribution of strings to the matter power
spectrum is suppressed by a factor of $36$ since the primordial string
perturbations are isocurvature instead of adiabatic.

To obtain the above result, we have assumed that the string-induced
fluctuations and the adiabatic perturbations are statistically
uncorrelated. If they were correlated, a cross term in the power
spectrum would appear which would only be suppressed by a factor
of $6$. We have also assumed that bias is unimportant, i.e.
that the observed galaxy correlation function equals the calculated
matter power spectrum. However, in a cosmic string model we expect
significant biasing (in particular in a model in which the
dark matter has a significant component of hot matter). However,
in the absence of a better understanding of the dynamics of cosmic
strings (in particular whether string loops \cite{TuB} or string wakes
\cite{SiVi,Va,StVe} are dominant), no firm conclusions can be
drawn.

Even if it turns out that the addition of a cosmic string component to
the primordial power spectrum does not change the cosmological neutrino
bounds, this addition will be important for many cosmological issues,
in particular for early structure formation \cite{Moessner} and
reionization \cite{Avelino}.

\vskip.5cm
\centerline{\bf Acknowledgments}

We are grateful to Guy Moore for interesting conversations at the beginning
of this project. R.B. acknowledges the hospitality of the
McGill Physics Department and the Fields Institute during the course
of this work. At Brown, this work was supported in part by the 
U.S. Department of Energy under Contract DE-FG02-91ER40688, TASK A. 
M.Y. is partially supported by the Japanese Grant-in-Aid for Scientific
Research from the Ministry of Culture, Sports, Science, and Technology.
A.M. acknowledges support from CITA and NSERC.

\end{document}